\documentclass[review]{elsarticle}

\usepackage{lineno,hyperref}
\usepackage{graphics}\usepackage{graphicx}
\usepackage{multirow}
\usepackage{caption}\usepackage{subcaption}
\graphicspath{ {./manuscript/} }
\modulolinenumbers[200]

\journal{Journal of New astronomy}

\usepackage{threeparttable}
\usepackage{tablefootnote}

\begin{document}

\begin{frontmatter}

\title{Global climate by Rossby number in the Solar system planets}
\author[mymainaddress]{Sora Lee}\corref{emailadress}
\ead{sora@naver.com}
%% Group authors per affiliation
\author[mymainaddress]{Maurice H.P.M. van Putten\corref{mycorrespondingauthor}}
\cortext[mycorrespondingauthor]{Corresponding author}
\ead{mvp@sejong.ac.kr}

\address[mymainaddress]{Room 614,Astronomy and Space Science, Sejong University, 98 Gunja-Dong Gwangin-gu,Seoul 143-147, Korea}

\begin{abstract}
On the largest scales, planetary climates can be described by their Rossby number (\textit{Ro}). \textit{Ro} is in response to $Gr/Re^2$ , where Gr is the Grashof number and Re is the Reynolds number. We here simplify $Gr/Re^2$ as h, where $h=H/H_{Earth}$ with $H=gP/(2\pi V_e)$ for a planet with surface gravity \textit{g}, rotation period \textit{P} and equatorial velocity V\textsubscript{e}. Unlike \textit{h}, \textit{Ro} is difficult to obtain because of a large diversity in observation. We perform on an in-depth literature search on average (av) and maximum (mx) wind velocity for each planet in the Solar system by various observational methods and by altitude. We explore a correlation between \textit{Ro} and \textit{h} expressed as a power law with index $\alpha$ based on wind velocities of planets in the Solar system. We obtain a correlation between \textit{Ro} and \textit{h} with $\alpha=0.56$ (av) and $\alpha=0.52$ (mx). Earth's $H=H_{Earth}$ ($h=1$) is primarily due to lunar tidal interaction, given our relatively distant habitable zone (HZ) to the Sun. Our positive correlation, therefore, suggests exoplanet-moon systems as the `go-to-place' in our searches for potentially advanced life in exosolar system.
\end{abstract}


\end{frontmatter}

%\linenumbers

\section{introduction}
Currently, there is a vigorous quest for potentially habitable exoplanets, following the pioneering discoveries by Sagittarius Window Eclipsing Extrasolar Planet Search (SWEEPS) in HST \cite{SWEEPS2009}, \textit{Convection, Rotation and planetary Transits} (CoRoT) \cite{CoRoT2009}, Kepler \cite{Kepler2010}, Kepler$'$ Second light (K2) \cite{K22014}, Gaia \cite{Gaia2008}, \textit{Arcsecond Space Telescope Enabling Research in Astrophysics} (ASTERIA) \cite{ASTERIA2017} and, today, \textit{Transiting Exoplanet Survey Satellite} (TESS) \cite{TESS2015}. 

At present, conditions of habitability remain somewhat speculate, though general conditions favoring liquid water and, perhaps, an oxygen-rich atmosphere are essential \cite{Franck2000}. Of interest also, and perhaps even more speculate, are conditions favoring advanced life. In particular, these conditions may not be the same as those for abiogenies, suggesting the need for co-evolution of life and planetary conditions as a whole. Here, we explore some conditions for a global clement climate relevant to the development of advanced life on Earth.
\par Planetary climates in our Solar system can be roughly divided into two groups: clement climates on Earth, Mars and Pluto and extreme climates on Jupiter, Saturn, Uranus and Neptune. We shall omit Mercury and Venus. Mercury is omitted for its essentially complete lack of atmosphere; Venus, tidally locked to the Sun, is heated one-sided driving a global climate different from those of the other planets in the Solar system. These kind of different wind patterns are commonly characterized by their Rossby number,
\begin{equation}
    Ro=\frac{U}{2\Omega R},
\end{equation}
where \textit{U} is the wind velocity of a planetary atmosphere, $\Omega$ is the angular frequency of planetary rotation. A global clement climate is proposed as a natural condition conducive for advanced life \cite{vanPutten2018}.
\par Quite generally, the global climate of a planet is a buoyancy-driven atmospheric flow on a rotating sphere subject to Coriolis forces. The driving force of buoyancy is governed by the Grashof number 
\begin{equation}
    Gr=\frac{\beta\Delta T R^3 g}{\nu^2}
\end{equation}
and the induced large scales flows are characterized by the Reynolds number,
\begin{equation}
   Re=\frac{UR}{\nu},
\end{equation}
where \textit{g} is surface gravity, $\beta$ is the coefficient of thermal expansion $\beta=1/T$, $\Delta T$ is a characteristic scale of the driving temperature difference (e.g. polar-to-equator) induced by exposure to the Sun, \textit{R} is the radius of a planet, $\nu$ is the kinematic viscosity, and $U=V_e$ is the equatorial velocity, $V_e=\Omega R$.

On the largest scales, the resulting buoyancy-driven flows are effectively described by $Gr/Re^2$. Here, ignoring variants of $\beta\Delta T$ across the different planets, we simplify 
\begin{equation}
    \frac{Gr}{Re^2} \approx \frac{gP}{2\pi V_e } \equiv H
\end{equation}
and normalize
\begin{equation}
    h=\frac{H}{H_{Earth}}. 
\end{equation}

\textit{h} was recently proposed as a habitability index for its general correlation to global climate \cite{vanPutten2018}. Here, we seek to quantify the correlation of \textit{Ro} to \textit{h} in the planetary atmospheres of our Solar system (\S3). \textit{Ro} is not so easy to infer because of a large diversity in observational measurements and associated uncertainties while \textit{h} is well defined for all Solar system planets. For this reason, we focused on an in-depth literature search on wind velocity data of all planets in the Solar system.
\par In \S2, we organize the wind velocities inferred for planets in the Solar system by observing methods and altitudes. In \S3, we report on a correlation of \textit{Ro} to \textit{h} using available data on average and maximum wind velocities of planets in the Solar system. In \S4, we summarize our exploration with a future outlook.

\newpage

\section{Data}
Wind velocities of planets in the Solar system are observed across different observing atmospheric heights (altitudes) by different methods ranging from ground-based, space-based and in-situ measurements. This poses a challenge in our effort to derive a homogeneous set of data.
 %Table 1-1.

\begin{table}[!ht]
    \centering
    \label{table:1.1}
    \renewcommand{\thetable}{1.\arabic{table}}
    \caption{Average wind velocity data for each planet in the Solar system by various observational methods.}

    \resizebox{\textwidth}{!}{
    \begin{tabular}{|l|l||c|c|c|c|c|c|c|}
    \hline
    \multicolumn{2}{|c|}{Methods}&Earth&Mars& Jupiter&Saturn&Uranus&Neptune&Pluto \\ 
    \hline\hline 
    \multirow{6}{*}{\rotatebox{90}{Image Analysis}}&Cloud& & & & &$84^{b}$&$325^{de}$& \\
    &Movement& & & & & & & \\
    \cline{2-9} 
    &Atmospheric& & &$100^{f}$&$400^{f}$& & & \\
    &Marking& & & & & & & \\
    \cline{2-9}
    &Surface& &55& & & & &$10^{h}$ \\
    &Topography& &$\pm21^{g\star}$& & & & & \\
    \hline 
    \multirow{6}{*}{\rotatebox{90}{Radio Observation}}&Doppler& & &90& & & & \\
    &shift& & &$\pm10^{l}$& & & & \\
    \cline{2-9}
    &Occultation& & & & &110& & \\
    & & & & & &$\pm40^n$ & & \\
    \cline{2-9}
    &Rotation& & & & &$200^{p}$&$400^{p}$& \\
    &period& & & & & & & \\
    \hline
    \multirow{4}{*}{\rotatebox{90}{In-situ Probes}}&Descent&50& & & & & & \\ 
    &Probes&$\pm10^{i}$& & & & & & \\
    \cline{2-9} 
    &\multirow{2}{*}{Landing}& &23& & & & & \\ 
    & & &$\pm4^{j}$& & & & & \\
    \hline 
    \multirow{2}{*}{\rotatebox{90}{Spectra}}&Thermal& & & & & & & \\
    &Emission& & & & & & & \\
    \hline
    \multicolumn{2}{|c|}{$^*$unspecified}& & &$100^{p}$&$500^{r}$& & & \\
    \hline\hline 
    \multicolumn{2}{|c|}{Average}&50&39&97&450&97&$^\bullet$325&10 \\
    \multicolumn{2}{|c|}{values}&$\pm10$&$\pm23$&$\pm6$&$\pm71$&$\pm18$& & \\
    \hline
\end{tabular}
}
\end{table}
\vspace*{-\baselineskip}

%\begin{tablenotes}
    \footnotesize
    \vskip0.1in
    \par $*$: Unspecified: cited papers did not mention methods.
    \par $\star$: 55 m/s is the average of 70 m/s, the wind velocity required for activation of Meridiani \par Planum observed by {\em Opportunity}, and 40 m/s, estimated from image of reversing dust \par  streaks (Mars Orbiter). 
    \par $\bullet$: The mean wind velocity of Neptune, 325 m/s, is measured also by Voyager 2 with no \par uncertainty reported.
    \par $\circ$: $85\pm7$ is westward jet velocity in both hemispheres, though a further 40 m/s eastward \par in the Southern hemisphere exists.
    \par a) Limaye, S. S., 2010 \cite{Limaye2010}; b) Hammel, H. B., 2001 \cite{Hammel2001}; c) Sromovsky, L. A. et al, 2005 \par \cite{Sromovsky2005}; d) Hammel, H. B. 1989 \cite{Hammel1989}; e) Smith, B. A., et al, 1989 \cite{Smith1989}; f) Hide, R. 1984 \cite{Hide1984}; \par g) Jerolmack, D. J. et al., 2006 \cite{Jerolmack2006}; h) Telfer, M. W. et al., 2018 \cite{Telfer2018}; i) Keegan, T. J. \par 1961 \cite{Keegan1961}; j) Pollack, J. B. et al., 1976 \cite{Pollack1976}; k) Santee, M. L. et al, 1995 \cite{Santee1995}; l) Atkinson, \par D. H. et al, 1998 \cite{Atkinson1998}; m) Tellmann, S. et al, 2013 \cite{Tellmann2013}; n) Lindal, G. F., 1987 \cite{Lindal1987}; o) \par Zalucha, A.M. 2016 \cite{Zalucha2016}; p) Helled, R. et al., 2010 \cite{Helled2010}; q) Guide to Space - Universe \par today \cite{Universetoday}; r) Ingersoll, A. P., 1990 \cite{Ingersoll1990}; s) Planets, Moons, and Dwarf Planets | NASA \par \cite{NASAwepsite}
%\end{tablenotes}

\vskip0.1in

Occasionally, we compromise, e.g., taking wind velocities from the mesosphere of Earth - with negative temperature gradient similar to that at low altitudes - alongside space-based observation of the other planets.

\par Observational techniques to measure wind velocities of a planet are mostly by image analysis, radio observations, in-situ probes and spectral methods (Tables 1.1-2). For instance, wind velocities can be obtained from images of cloud movement, atmospheric markings, e. g., Jupiter$'$s Giant Red Spot, and surface topography, e.g., Pluto$'$s dunes. Also, wind velocities can be obtained via radio observation: radio Doppler shift, radio occultation, and radio rotation period. Additionally, wind velocities have been measured by in-situ probes: descent probes and Landers. Furthermore, wind velocities can be obtained through spectral methods such as thermal emission. Tables 1.1-2 summarize our literature search on wind velocities for each planet in the Solar system by various methods. In this table, uncertainties refer to scatter in data from multiple measurements reported in the literature. For this broad range of measurement approaches, our references are listed in the notes of Table 1.1-2.
%Table 1-2
\begin{table}[!ht]
    \centering
    \label{table:1.2}
    \renewcommand{\thetable}{1.\arabic{table}}
    \caption{Maximum wind velocity data for each planet in the Solar system by various observational methods.}

    \resizebox{\textwidth}{!}{
    \begin{tabular}{|l|l||c|c|c|c|c|c|c|}
    \hline
    \multicolumn{2}{|c|}{Methods}&Earth&Mars& Jupiter&Saturn&Uranus&Neptune&Pluto\\ 
    \hline\hline
    \multirow{6}{*}{\rotatebox{90}{Image Analysis}}&Cloud&$50^{a}$&$40^{a}$&$170^{a}$&$450^{a}$&240&$560^{e}$& \\
    &Movement& & & & &$\pm50^{c}$& & \\ 
    \cline{2-9}
    &Atmospheric& & & & & & &$10^{h}$ \\
    &Marking& & & & & & & \\
    \cline{2-9}
    &Surface& & & & & & & \\
    &Topography& & & & & & & \\ 
    \hline
    \multirow{6}{*}{\rotatebox{90}{Radio Observation}}&Doppler& & &175& & & & \\
    &shift& & &$\pm25^{l}$& & & & \\ 
    \cline{2-9}
    &Occul& &$170^{m}$& & & & &$10^{o}$ \\
    &tation& & & & & & & \\
    \cline{2-9}
    &Rotation& & & & & & & \\
    &period& & & & & & & \\
    \hline
    \multirow{2}{*}{\rotatebox{90}{In-situ Probes}}&Descent&$70^{i}$& & & & & &  \\ 
    &Probes& & & & & & & \\ 
    \cline{2-9}
    &\multirow{2}{*}{Landing}& & & & & & & \\
    & & & & & & & & \\
    \hline
    \multirow{2}{*}{\rotatebox{90}{Spectra}}&Thermal& &85& & & & & \\
    &Emission& &$\pm7^{k\circ}$& & & & & \\
    \hline
    \multicolumn{2}{|c|}{unspecified}& & &$172^{q}$&$500^{q}$&$250^{qs}$&$556^{s}$& \\ 
    \hline\hline
    \multicolumn{2}{|c|}{Maximum}&60&98&172&475&235&505&10 \\
    \multicolumn{2}{|c|}{value}&$\pm14$&$\pm66$&$\pm3$&$\pm35$&$\pm24$&$\pm91$& \\ 
    \hline

\end{tabular}
}
\end{table}

\par Pluto is an interesting special case with wind velocities estimated for the first time by the \textit{New Horizons} mission \cite{Horizons2008}. In its extremely tenuous atmosphere, wind velocities are now inferred from atmospheric temperature structure and also from dunes on its surface. The first shows that winds concentrate about polar regions, from greater than 10 m/s down to a few m/s at intermediate latitudes \cite{Zalucha2016}. This is corroborated by wind velocities of about 10 m/s inferred from dunes at similar intermediate latitudes \cite{Telfer2018}. Thus, we use Pluto$'$s wind velocity of 10 m/s in Table 1.1-2. With no known uncertainties, uncertainty is left unspecified in Table 1.1-2.
\vskip0.2in
\par Quite generally, wind velocities vary with altitude. As a compromise in deriving a reasonably homogeneous data set, we would focus on velocities in layers with negative temperature gradients. On Earth, there are two such layers, at low altitudes 0-10 km and at high altitudes 50-80 km. Most other planets in the Solar system have a similar structure, excluding perhaps Uranus and Mars \cite{Andrew2008}. In Table 2, we list results for Earth, Mars and Pluto by altitude. These data are indeed consistent with the mean of the average (av) and maximum (ma) values of Table 1.1-2. On this admittedly limited basis, we shall proceed with the data for all planets in Table 1.1-2 despite the absence of such explicit data by altitude for the other four planets.
%Table 2
\begin{table}[!ht]
    \centering
    \label{table:2}
    \renewcommand{\thetable}{2}
    \caption{Wind velocities of the planets in the Solar system in the two temperature-negative gradient regions by altitude. See the notes of Table 1.1 for references.}
    \begin{tabular}{||c|c|c|c||} 
    \hline
     Altitude(km) & Earth & Mars & Pluto \\ [0.5ex] 
    \hline\hline
    $50\sim80$ & $50\pm10^{ac}$ & $85\pm5^{r}$ &  \\ 
    \hline
    $0\sim10$ &  & $40^{a}$ & $10^{g}$ \\ [1ex] 
    \hline
\end{tabular}
\end{table}

\section{A \textit{Ro}-\textit{h} correlation}
We calculate \textit{Ro} from the wind velocities of Table 1.1-2 and \textit{h} from the physical parameters of planets in the Solar system. Table 3.1-2 summarizes our estimates of \textit{Ro} and \textit{h} for all planets in the Solar system excluding Mercury and Venus.
%Table 3
\begin{table}[!ht]
    \centering
    \label{table:3.1}
    \renewcommand{\thetable}{3.1}
    \caption{Average values of Ro and \textit{h} for all planets in the Solar system, where $\hat Ro=Ro/Ro_{Earth}$.}
    \begin{tabular}{|c||c|c|c|c|c|c|c|} 
    \hline
    &Earth&Mars&Jupiter&Saturn&Uranus&Neptune&Pluto\\ [0.5ex] \hline\hline
    \textit{Ro}&95.2&144&6.8&38.9&33.2&106&651\\ 
    \hline
    $\hat Ro$&1&1.51&0.07&0.41&0.35&1.11&6.84\\ 
    \hline
     \textit{h}&1.00&0.76&0.04&0.02&0.12&0.14&13\\ [1ex]
    \hline
\end{tabular}
\end{table}

\begin{table}[!ht]
    \centering
    \label{table:3.2}
    \renewcommand{\thetable}{3.2}
    \caption{Maximum values of \textit{Ro} and \textit{h} for all planets in the Solar system, where $\hat Ro=Ro/Ro_{Earth}$.}
    \begin{tabular}{|c||c|c|c|c|c|c|c|} 
    \hline
    &Earth&Mars&Jupiter&Saturn&Uranus&Neptune&Pluto\\ [0.5ex] \hline\hline
    \textit{Ro}&114&362&12&41.1&80.4&164&651\\ \hline
    $\hat Ro$&1&3.17&0.1&0.36&0.70&1.44&5.70\\ \hline
    \textit{h}&1.00&0.76&0.04&0.02&0.12&0.14&13\\ [1ex]
    \hline
\end{tabular}
\end{table}

\par We consider the correlation of \textit{Ro} to \textit{h} be of the form of a power law:
\begin{equation}
    Ro=c_{0}\,\left(\frac{Gr}{Re^2}\right)^\alpha \propto~ h^\alpha,
\end{equation}
\par To estimate the power law index $\alpha$, we draw a loglog plot of our \textit{Ro} to \textit{h} and calculate $\alpha$ by using the MatLab \textit{fitnlm} (\textit{h}, \textit{Ro}, \textit{y}, $B_0$). Here, \textit{y} is the linear function $y=ax+b$ with $y=\log_{10} Ro $ and $x=\log_{10} \textit{h}$ ,and $B_0$ are initial data of the unknown coefficients \textit{a} and \textit{b}.  
\par Figure 1 shows our result for Ro using average wind velocities and our estimate $\alpha=0.56$ with standard error of 0.15, giving equal weights to all planets in the Solar system. The same figure shows our result for Ro using maximum wind velocities and our estimate $\alpha=0.52$ with standard error of 0.14 obtained similarly.

\begin{figure}[!ht]
    \centering
    \begin{subfigure}[t]{0.45\textwidth}
    \centering
    \includegraphics[width=\linewidth]{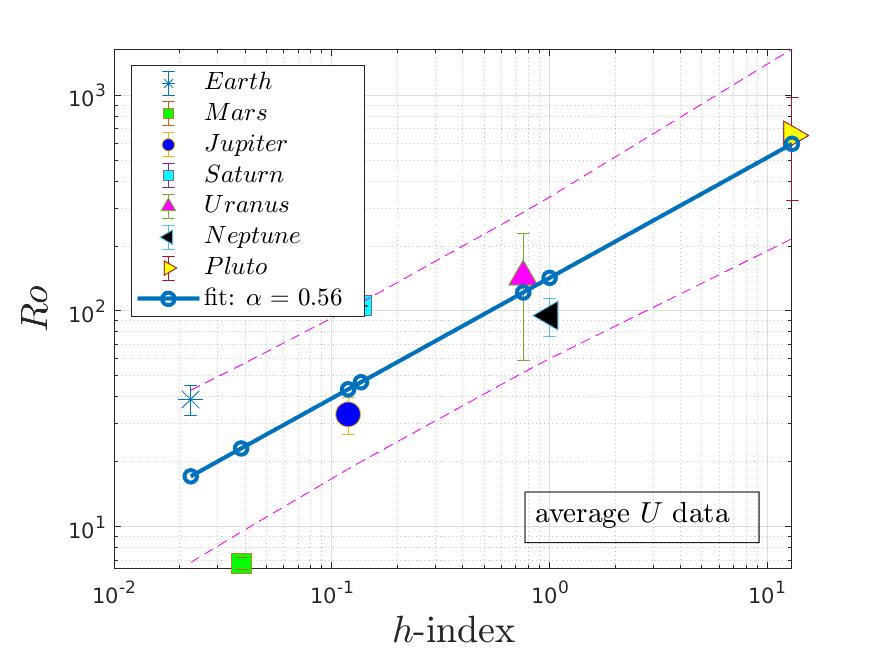}
    \end{subfigure}
    \hfill
    \begin{subfigure}[t]{0.45\textwidth}
    \includegraphics[width=\linewidth]{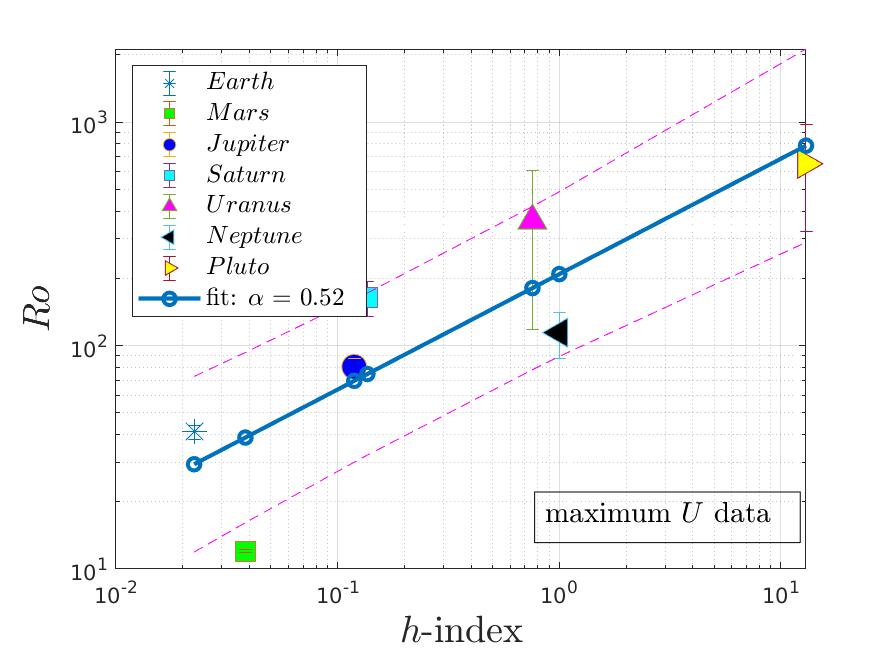}
    \end{subfigure}
    \caption{The power law correlation of \textit{Ro} and \textit{h} using average wind velocities (left panel) and maximum wind velocities (right panel) based on equal weights to all planets in the Matlab routine \textit{fitnlm} with $\pm1\sigma$ (pink line)\label{fig:1.1}}
\end{figure}

\section{Conclusion and Outlook}
Here, we have reviewed wind velocity data of the planets in the solar system by observational methods (Table 1.1) and altitudes (Table 1.2). We used this review to explore a novel correlation of \textit{Ro} and \textit{h}. 
\par \textit{Ro} characterizes the response to $Gr/Re^2$ measured by wind velocity. Jupiter, Saturn, Uranus and Neptune have low \textit{Ro} and indeed show global extreme climate. On the other hand, Mars, Earth and Pluto have high \textit{Ro} and indeed show global clement climate. 
\par \textit{h} is our reduced expression for $Gr/Re^2$ and, relatively easy to observe for planets of our Solar system, may indeed indicate potentially favorable conditions to advanced life. Jupiter, Saturn, Uranus and Neptune have small \textit{h}-values because they are fast spinning whereas Mars, Earth and Pluto are slow spinning, indicated by the respective \textit{h}-values.
\par Quantitatively, \textit{Ro} and \textit{h} satisfy a power law with index of $\alpha =0.56$ and $\alpha=0.52$ based on average (av) and, respectively, maximum (mx) wind velocities in Table 1.1-2, upon attributing equal weights to all planets.
\par While the observed positive correlation of \textit{Ro} and \textit{h} appears to be robust, there is clearly a need for data with a much more uniform uncertainty than which is available today, especially by what would be an anomalous tilt by Jupiter in estimates of $\alpha$ weighted by present uncertainties.
\par However, \textit{h} will be difficult to observe for relatively small potentially Earth-like exoplanets. Quite generally, \textit{h} of order unity corresponds to slow spin. Slow spin, in turn, can be attributed to lunar tidal interactions in relatively distant habitable zone‘s such as ours around the Sun. It appears, therefore, that exoplanet-moon systems are possibly the $“$go-to-places$”$ for advanced life in exosolar systems \cite{vanPutten2017}.
\par The upcoming James Webb Space Telescope (JWST) \cite{JWST2006} and the Extremely Large Telescope (ELT) \cite{ELT2011} may guide us to such systems in the near future.

\section*{Acknowledgment}
The authors thank the reviewer for a detailed reading of the
manuscript and constructive comments.
Support is acknowledged from the National Research Foundation of Korea under grants 2015R1D1A1A01059793, 2016R1A5A1013277 and 2018044640.

\end{document}